\documentclass[aps,prl,reprint,notitlepage,showpacs,nofootinbib,superscriptaddress,twocolumn
,showkeys,preprintnumbers,floatfix]{revtex4-1}
\usepackage{graphicx}
\usepackage[utf8]{inputenc} 
\usepackage{amsmath}
\usepackage{amssymb}
\usepackage{ulem}
\usepackage{comment}
\usepackage{slashed}
\usepackage{subfigure}

\usepackage[T1]{fontenc} 
\usepackage{parskip}
\usepackage{bm}         
\usepackage{array}
\usepackage{multirow}   
\usepackage{booktabs}   

\usepackage{textcomp}
\usepackage{amssymb}
\usepackage{amsmath}
\usepackage{soul}

\usepackage{enumerate}
\usepackage{ragged2e}

\usepackage{amsmath}
\usepackage{amssymb}
\usepackage{indentfirst}
\usepackage[dvipsnames]{xcolor}
\usepackage{silence}
\usepackage[colorlinks=true,citecolor=darkred,urlcolor=darkred, pdfborder={0 0 0}]{hyperref}
\definecolor{darkred}{rgb}{0.6,0,0}

\definecolor{linkcolor}{rgb}{0,0,0.5}
 
\newcommand {\ignore}[1]{}

\definecolor{bostonuniversityred}{rgb}{0.8, 0.0, 0.0}

\usepackage{soul}
\definecolor{mightnightblue}{RGB}{25,25,112}

\definecolor{brown}{rgb}{0.59, 0.29, 0.0}

\def\21{$\mathrm{SU(2)_L \otimes U(1)_Y}$}

\newcommand{\AddrUNAM}{ {\it Instituto de Física, Universidad Nacional Autónoma de México
Ciudad de México, C.P. 04510, Mexico}}

\newcommand{\AddrLJF}{ {\it Laboratorio Interdisciplinario de Investigación Tecnológica-LIIT, Tecnol\'ogico Nacional de M\'exico/ITS de Jerez, C.P. 99863, Zacatecas, M\'exico.}}

\newcommand{\AddrFC}{ {\it Departamento de Física, Facultad de Ciencias, Universidad Nacional Autónoma de México,
Apartado Postal 70-542, Ciudad de México 04510, Mexico.}}

\begin{document}

\title{Constraints on Light Sterile Neutrinos and Scalar Non-Standard Interactions Using the First Reactor Antineutrino Oscillation Results at JUNO}

\author{L. J. Flores}\email{ljflores@jerez.tecnm.mx }
\affiliation{\AddrLJF}

\author{R.~Pacheco-Ak\'e}\email{rodrigopa@estudiantes.fisica.unam.mx }
\affiliation{\AddrUNAM}

\author{Eduardo Peinado}\email{epeinado@fisica.unam.mx}
\affiliation{\AddrUNAM}

\author{G.~Sanchez Garcia}\email{g.sanchez@ciencias.unam.mx }
\affiliation{\AddrFC}

\author{E.~V\'azquez-J\'auregui}\email{ericvj@fisica.unam.mx }
\affiliation{\AddrUNAM}

\vspace{0.7cm}

\begin{abstract}

Constraints on light sterile neutrinos and scalar non-standard neutrino interactions are obtained from the first reactor antineutrino results reported by JUNO. The analysis is based on a spectral $\chi^2$ fit to the prompt-energy distribution corresponding to 59.1 days of data, including full three-flavor oscillations extended to a $3+1$ framework and effective scalar NSI contributions. The reactor flux is modeled using the Daya Bay measured spectrum, and systematic uncertainties are accounted for through a set of nuisance parameters describing reactor flux normalization, spectral shape, background normalization, and detector response. It is found that JUNO is already sensitive to light sterile neutrinos in the mass-splitting range $10^{-5} \lesssim \Delta m^2_{41}/\text{eV}^2 \lesssim 10^{-2}$, probing mixing amplitudes down to $\sin^2 2\theta_{14} \sim \mathcal{O}(10^{-1})$. In addition, a constraint on the scalar NSI parameter $|\eta_{ee}| \sim \mathcal{O}(10^{-3})$ is obtained, with correlations with solar oscillation parameters. These results demonstrate the potential of JUNO to probe small deviations from the Standard Model resulting from new physics through precision measurements, with significant improvements expected as statistics and systematic control improve.

\end{abstract}

\maketitle

\section{Introduction}

The discovery of neutrino oscillations established that neutrinos are massive particles and mix, providing the first confirmed evidence of physics beyond the Standard Model (SM)~\cite{Fukuda:1998mi,Ahmad:2002jz,Eguchi:2002dm}. In the standard three-flavor framework, neutrino propagation is governed by the unitary Pontecorvo–Maki–Nakagawa–Sakata (PMNS) matrix~\cite{Pontecorvo:1957cp,Maki:1962mu}. Precision measurements of oscillation parameters by solar, atmospheric, reactor, and accelerator experiments~\cite{KamLAND:2008dgz,SuperK:2017yvm,T2K:2020vkk,NOvA:2021CDR,DayaBay:2022orm,PDG:2024neutrino} have established this framework with high accuracy. Nevertheless, high-precision measurements provide a powerful probe of new physics, including sterile neutrinos and non-standard neutrino interactions (NSI).

Reactor antineutrino experiments play a central role in precision tests of neutrino physics. Their advantages include an intense $\bar{\nu}_e$ flux and sensitivity to oscillation frequencies through spectral distortions. Past experiments such as KamLAND~\cite{KamLAND:2008dgz}, Daya Bay~\cite{DayaBay:2022orm}, RENO~\cite{RENO:2018pdp}, and Double Chooz~\cite{DoubleChooz:2019qbj} have provided stringent constraints on the absolute flux, $\theta_{13}$, and $\Delta m^2_{21}$ while enabling tests of new physics including sterile neutrinos and NSI.

The Jiangmen Underground Neutrino Observatory (JUNO)~\cite{JUNO:2015zny} is the next generation of reactor antineutrino experiments. With a 20~kt liquid scintillator target, an average of 52.5~km baseline to all the reactor cores of the Taishan and Yangjiang reactors, and a design energy resolution of $3\%$ at $1~\mathrm{MeV}$, JUNO is designed to resolve the interference pattern between oscillation frequencies at an unprecedented level. The experiment’s primary goals include determining the neutrino mass ordering and achieving subpercent-level precision on the oscillation parameters~\cite{JUNO:2022mxj}. JUNO is also sensitive to explore physics Beyond the Standard Model, such as light sterile neutrinos and scalar NSI.

JUNO is sensitive to light sterile neutrinos with mass splittings in the range $\Delta m_{41}^2 \sim 10^{-5}$--$10^{-2}\,\mathrm{eV}^2$. This sensitivity arises from the large baseline of $L \simeq 52.5$~km combined with the excellent energy resolution of the detector, which allows the experiment to resolve slow oscillation patterns in the reactor antineutrino spectrum. In contrast, shorter baseline experiments such as Daya Bay ($L \sim 1$--$2$~km) lose sensitivity in this region because the oscillation phase ($1.27\,\Delta m_{41}^2 L/E$) becomes too small to generate observable spectral distortions. As a result, JUNO is able to probe sterile-neutrino oscillations in a region that is largely inaccessible to short-baseline reactor experiments~\cite{Girardi:2014gna}. This behavior follows from the oscillation phase, which shows that longer baselines enhance sensitivity to small mass splittings by increasing the oscillation phase at fixed neutrino energy.

In addition to sterile neutrinos, reactor experiments can also probe non-standard neutrino interactions, including scenarios mediated by new scalar particles. In such scenarios, scalar couplings can modify the effective neutrino mass matrix and induce distortions in the measured oscillation spectrum. These effects can be parametrized through effective couplings such as $\eta_{ee}$, which alter the electron-flavor component of the neutrino Hamiltonian and introduce correlations with the determination of the solar oscillation parameters. Precision reactor experiments therefore provide a way to test these effects through detailed measurements of the antineutrino spectrum. General frameworks for non-standard neutrino interactions have been extensively studied~\cite{Wolfenstein:1977ue,Ohlsson:2012kf,Miranda:2015dra, Farzan:2017xzy}, while scalar-mediated neutrino interactions and their impact on neutrino masses and oscillations have also been explored~\cite{Ge:2018uhz,Babu:2019iml,Dev:2020kgz,Gupta_2025, Denton:2024upc,Choubey:2026jiq, Singha:2023, Jana_2025, dutta2025, Denton_2023}.

In this work, the recently released JUNO results~\cite{JUNO:2025results} are analyzed to derive new constraints on light sterile neutrinos and scalar NSI. The reactor antineutrino spectrum is modeled using the Daya Bay measured spectrum~\cite{An:2025dbspectrum} and normalized to an antineutrino flux of $4.22\times10^{6}~\bar{\nu}_e/(\mathrm{cm}^2\mathrm{s})$. This normalization corresponds to the expected reactor antineutrino flux at JUNO for a total reactor thermal power of 26.6~GW$_{\rm th}$ and an average baseline of 52.5~km from the reactor cores, consistent with the normalization reported in the JUNO oscillation analysis~\cite{JUNO:2025results}. The predicted event rates incorporate full three-flavor oscillations, and the contributions of sterile neutrinos or NSI. A $\chi^2$ fit to the measured energy spectrum of prompt inverse beta decay (IBD) candidates is performed, including systematic uncertainties associated with the reactor flux normalization, spectral shape, background normalization, and detector energy response, 
including energy scale, resolution, and non-linearity.

\section{Theoretical Framework}

\subsection{Light Sterile Neutrinos}

The current understanding of neutrino physics involves oscillations between three different active states. This picture is robust and most of its parameters are well determined~\cite{deSalas:2020pgw,Esteban:2024eli,Capozzi:2025wyn}. However, different theoretical motivations and experimental results point towards the possibility of the existence of a fourth light sterile neutrino~\cite{Acero:2022wqg}. Sterile neutrinos are theorized singlets under the Standard Model gauge symmetries. They are called \textit{sterile} in the sense that they do not take part
in weak interactions. However, they can mix with the active neutrinos and therefore induce oscillation from the active to the sterile one.

While there is not a limit on the allowed number of sterile neutrinos, nor their mass scale, there are existing bounds on the active-sterile mixing parameters assuming one additional light state. The existence of sterile neutrinos has been investigated at several different scales. For instance, sterile neutrinos in the range close to the GUT scale can explain the smallness of active state masses,  through the so-called Type-I seesaw mechanism~\cite{Minkowski:1977sc,Yanagida:1979as,Glashow:1979nm,Gell-Mann:1979vob,Mohapatra:1980yp,Schechter:1980gr}, while those within the keV scale, are candidates for warm dark matter~\cite{Kusenko:2009up}. On the other hand, those in the eV range were originally studied as a solution to the $\bar{\nu}_e$ excess observed by the LSND~\cite{LSND:2001aii} and the MiniBooNE~\cite{MiniBooNE:2010idf} experiments, as well as the disappearance anomaly of $\nu_e$ observed during the calibration of solar experiments like GALLEX and SAGE~\cite{Giunti:2010zu}. However, the recent results by the MicroBooNE collaboration~\cite{MicroBooNE:2025nll} disfavored the single sterile neutrino interpretation of the LSND and MiniBooNE anomalies and a large portion of the parameter space available to explain the gallium anomalies. Other studies have also constrained the sterile neutrino flux from solar neutrino experiments~\cite{Ankush2019}. Nonetheless, light sterile neutrinos are still not completely ruled out and can be investigated in long-baseline experiments like JUNO.

Here, a $3+1$ framework is adopted, consisting of the three active neutrinos and one light sterile state. The $4\times4$ matrix $\mathcal{U}$ relates the flavor and mass basis. Hence, the usual survival probability of electron antineutrinos produced at reactor facilities will be modified, giving the form~\cite{JUNO:2015zny}

\begin{widetext}
\begin{equation}\begin{array}{lcl}
P_{\bar\nu_e\to\bar\nu_e}^{\mathrm{S}} &=&  1
 - \cos^4 \theta_{14} \cos^4 \theta_{13} \sin^2 2\theta_{12} \sin^2 \Delta_{21}
 - \cos^4 \theta_{14} \sin^2 2\theta_{13} \Bigl(\cos^2 \theta_{12} \sin^2 \Delta_{31} + \sin^2\theta_{12} \sin^2 \Delta_{32}\Bigr) \\&&
 - \cos^4\theta_{13} \sin^2 2\theta_{14} \Bigl(\cos^2 \theta_{12} \sin^2 \Delta_{41}+ \sin^2\theta_{12} \sin^2 \Delta_{42}\Bigr)
 - \sin^2\theta_{13} \sin^2 2\theta_{14} \sin^2 \Delta_{43},
 \end{array}
\end{equation}
\end{widetext}
where $\theta_{ij}$ are the mixing angles and $\Delta_{ij} = \Delta m^2 _{ij} L / 4E$, with $ \Delta m^2 _{ij} = m_i^2 - m_j^2$, $L$ the distance from source to detector, and $E$ the incoming neutrino energy. In the following sections, this survival probability is used to test the capability of JUNO to constrain the oscillation parameters in the $3+1$ scheme.

\subsection{Non-Standard Interactions}

The mechanism responsible for neutrino masses remains an open question in modern physics. Several mass models have been proposed to explain their origin ~\cite{Schechter:1980gr, King:2003jb, Ma:2006km}. Although testing each model individually can be challenging, a phenomenological approach is often preferred, as it allows the effects of a broad class of models to be encoded within a common parametrization. For example, some models predict corrections to the vector and axial couplings of the SM. These corrections can be parametrized within the NSI formalism, represented by an effective Lagrangian with a structure similar to that of the SM, but allowing non-universal interactions and flavor transitions. In this framework, the strength of the interaction is encoded in NSI parameters, which can be tested by different experiments. Bounds on these NSI parameters can then be used to constrain the parameter space of specific models.

Neutrino interactions beyond the SM are not limited to vector and axial couplings. In fact, scalar and tensor interactions can also arise when massive Dirac neutrinos are considered~\cite{Bischer:2019ttk, Flore:2026qeh}. Here, the analysis focuses on the particular case where the new physics arises from the coupling to a scalar field, $\phi$, with mass $m_\phi$. Then, the effective Lagrangian parametrizing this interaction can be written as

\begin{equation*}
\mathcal{L}_{\text{sNSI}}^{\text{eff}}
=
\frac{y_f\, Y_{\alpha\beta}}{m_\phi^2}
\left[\bar{\nu}_{\alpha}(p_3)\nu_{\beta}(p_2)\right]
\left[\bar{f}(p_1)f(p_4)\right],
\end{equation*}
where $y_f$ and $Y_{\alpha\beta}$ are the Yukawa couplings of the fermion field with the matter fields $f = e, u, d$, and with neutrinos, respectively. Unlike vector NSI, which produces changes in the matter potential, this interaction induces modifications to the neutrino mass matrix. As a result, the effective Hamiltonian for neutrino propagation can be expressed as
\begin{equation}
\begin{aligned}
\mathcal{H}^{\text{eff}}_\mathrm{sNSI} \approx \frac{1}{2E}\Bigl[&\,(\mathcal{M}+\delta M)(\mathcal{M}+\delta M)^{\dagger}\\
&+2E\,V_\mathrm{CC}
\begin{pmatrix}
1 & 0 & 0 \\
0 & 0 & 0 \\
0 & 0 & 0
\end{pmatrix}\Bigr],
\end{aligned}
\end{equation}
where $\delta M = \sum_f \frac{N_f y_f Y_{}\alpha\beta}{m_\phi^2}$ encodes the scalar NSI contribution, which is proportional to the fermion density of the medium, $N_f$. In general, $\delta M$ can be parametrized as~\cite{Ge:2018uhz}
\begin{equation}
\delta M =
\sqrt{\left|\Delta m_{31}^{2}\right|}
\begin{pmatrix}
\eta_{ee} & \eta_{e\mu} & \eta_{e\tau} \\
\eta_{e\mu}^{*} & \eta_{\mu\mu} & \eta_{\mu\tau} \\
\eta_{e\tau}^{*} & \eta_{\mu\tau}^{*} & \eta_{\tau\tau}
\end{pmatrix},
\end{equation}
where $\eta_{\alpha\beta}$ are the scalar NSI strength parameters. For simplicity, only diagonal scalar NSI parameters are considered here. Therefore, only $\eta_{ee}$ is relevant for reactor antineutrino studies.

Since the presence of scalar NSI modifies the effective oscillation parameters, the electron antineutrino survival probability for the JUNO baseline keeps the standard form but with modified parameters~\cite{Gupta_2025}
\begin{equation}
\begin{aligned}
P^\mathrm{sNSI}_{\bar\nu_e\to\bar\nu_e} =
&\; 1
- \cos^{4}\theta_{13}\,\sin^{2}(2\theta_{12}^\prime)\,\sin^{2}\Delta_{21}^\prime \\
&- \sin^{2}(2\theta_{13})
\left[
\cos^{2}\theta_{12}^\prime\,\sin^{2}\Delta_{31}^\prime
\right.\\
&\left.\qquad + \sin^{2}\theta_{12}^\prime\,\sin^{2}\Delta_{32}^\prime
\right].
\end{aligned}
\end{equation}
where $\Delta_{ij}^\prime = \Delta m^{2\,\prime}_{ij}L /4 E$. The modified solar mixing angle and mass splittings are given by 
\[\tan 2\theta_{12}' =
\frac{\sin 2\theta_{12}\left(\Delta m_{21}^{2} + \epsilon_{3}\cos^{2}\theta_{13}\right)}
{\Delta m_{21}^{2}\cos 2\theta_{12} - \epsilon_{2}},\]
\[
  \begin{aligned}
\Delta m_{21}^{2\,\prime} =
&\;\Delta m_{21}^{2}\cos 2(\theta_{12}-\theta_{12}')  \\
&+ \cos^{2}\theta_{13}\sin 2\theta_{12}\sin 2\theta_{12}'\,\epsilon_{3}
- \cos 2\theta_{12}'\,\epsilon_{2},
\end{aligned}
\]
and
\[\begin{aligned}
\Delta m_{31}^{2\,\prime} =
&\;\Delta m_{31}^{2}
- \Delta m_{21}^{2}\sin^{2}(\theta_{12}-\theta_{12}') \\
&- \cos^{2}\theta_{12}'\,\epsilon_{2}
- \frac{1}{2}\cos^{2}\theta_{13}\sin 2\theta_{12}\sin 2\theta_{12}'\,\epsilon_{3}.
\end{aligned}
\]
Corrections on $\theta_{13}$ due to $\eta_{ee}$ are negligible. The dependence on the scalar NSI parameter $\eta_{ee}$ is encoded in $\epsilon_2$ and $\epsilon_3$, which, for normal mass ordering (NO) take the form 
$$\epsilon_2 = 2\sqrt{|\Delta m^2_{31}|} \,\eta_{ee}\, c_{13}^2 m_1 \left(c_{12}^2 + \frac{m_2}{m_1}s_{12}^2\right),$$ 
and $$\epsilon_3 = \sqrt{|\Delta m^2_{31}|}\, \eta_{ee} m_1 \left(-1 + \frac{m_2}{m_1}\right).$$ A benchmark value of $m_1 = 10^{-3}$ eV is adopted, consistent with the NO regime and with existing bounds.

An important aspect of scalar NSI is that their effects depend on the matter density, since the interaction comes from couplings with background fermions. Because of this, the oscillation parameters measured in terrestrial experiments are, in general, effective parameters that may already include contributions from scalar NSI. This leads to degeneracies between the standard oscillation parameters and the scalar NSI parameters, especially in experiments like JUNO, where the sensitivity mainly comes from the precise measurement of the solar parameters. To disentangle these effects, it is necessary to combine measurements taken at different matter densities, so that the impact of scalar NSI can be compared in different environments.

Here, it is assumed that scalar NSI are present. The analysis is performed within this framework, and constraints are obtained on the effective parameter $\eta_{ee}$. The correlations with $\sin^2\theta_{12}$ and $\Delta m^2_{21}$ that appear in the results are a consequence of this degeneracy.

\section{Experiment and Analysis Description}

The JUNO experiment is a multipurpose liquid scintillator detector located in Jiangmen, China, designed to measure electron antineutrinos produced by the Taishan and Yangjiang nuclear power plants. The combined reactor complex has a nominal thermal power of \(26.6~\mathrm{GW_{th}}\). The detector sits at an average baseline of 52.5~km and is shielded by approximately 700~m of rock overburden, significantly reducing cosmogenic backgrounds.

The central detector consists of 20~kt of linear-alkylbenzene (LAB) based liquid scintillator viewed by 17{,}596 large (20-inch) high-quantum-efficiency photomultiplier tubes (PMTs) and 25{,}587 small (3-inch) PMTs, providing excellent energy resolution and nearly isotropic light collection. Antineutrinos are detected via the inverse beta decay process $\bar{\nu}_e + p \rightarrow e^+ + n$. The prompt positron signal is followed by a delayed neutron-capture signal, predominantly on hydrogen, with a mean capture time of $\sim$200~$\mu$s. Surrounding muon veto systems and the water Cherenkov detector provide identification and rejection of cosmogenic background events~\cite{JUNO:2025results}.

This analysis uses the JUNO data presented in Ref.~\cite{JUNO:2025results}, corresponding to a livetime of 59.1~days. The absolute energy scale is calibrated to within 1\% using a combination of deployed radioactive sources and cosmogenic isotopes, and the detector achieved an energy resolution of approximately $3.4\%$. The function employed to model the energy resolution is 
\[\frac{\sigma_E}{E} = \sqrt{\frac{a^2}{E} + b^2},\]
where $a$ is found to be $\sim 3.3\%$ and the constant term $b$ is $\sim 1\%$~\cite{JUNO:2025fpc}. The background model employed here includes only the three dominant contributions reported in Ref.~\cite{JUNO:2025results}: the \(^{9}\mathrm{Li}/^{8}\mathrm{He}\) cosmogenic background, geoneutrinos, and antineutrinos from worldwide reactors. Their spectral shapes and nominal normalizations are taken directly from the JUNO results. Although only these three components are included explicitly in the spectral fit, the full set of background-related systematics listed in Table~1 of Ref.~\cite{JUNO:2025results} is incorporated through a single nuisance parameter accounting for the combined background uncertainty.

The predicted event rate at the detector is
\[
  \begin{aligned}
N(E_{\rm prompt}) &\propto P_{ee}(E)\,\Phi_{\bar{\nu}_e}(E)\\
&\quad\times\sigma_{\rm IBD}(E)\,R(E_{\rm prompt},E),
\end{aligned}
\]
where $\sigma_{\rm IBD}$ is the IBD cross-section~\cite{Vogel:1999zy, tomalak2025theoryinversebetadecay, tomalak2025radiativecorrectionsinversebeta}. The detector response function $R(E_{\rm prompt},E)$ is constructed from the measured energy scale and resolution.

The reactor antineutrino flux $\Phi_{\bar{\nu}_e}(E)$ is modeled using the 
Daya\,Bay measured spectrum~\cite{An:2025dbspectrum}, and normalized to an antineutrino flux of $4.22\times10^{6}~\bar{\nu}_e/(\mathrm{cm}^2\mathrm{s})$, consistent with JUNO flux expectations~\cite{JUNO:2025results}. Neutrino propagation is computed assuming a constant matter density of $2.6~\text{g/cm}^3$. Six nuisance parameters are included in the fit: $\alpha$, $\beta$, and $\gamma$, together with three additional parameters describing the detector response, $\delta_1$, $\delta_2$, and $\delta_3$. The parameter $\alpha$ mostly accounts for reactor-related normalization uncertainties affecting the predicted IBD rate, including the Daya\,Bay normalization uncertainty. Additionally, it includes a contribution from the uncertainty of the target proton. In this analysis, a prior uncertainty of $2.4\%$ is assigned to $\alpha$. The parameter $\beta$ describes the combined background uncertainty, for which a prior uncertainty of $20\%$ is adopted. The parameter $\gamma$ modifies the predicted signal spectrum to account for uncertainties in the shape of the reactor antineutrino spectrum:
\[
  S_i(\gamma) \;\longrightarrow\; S_i\,\left[1 + \gamma\bigl(E_i - \langle E\rangle\bigr)\right],\]
where $S_i$ denotes the predicted IBD signal, with $\langle E\rangle = 3.5~\mathrm{MeV}$ and a conservative prior uncertainty 
$\sigma_\gamma = 0.02$, corresponding to the shape uncertainty.
The parameters $\delta_k$ ($k=1,2,3$) are included to account for residual uncertainties in the detector energy response: 
$\delta_1$ models the absolute energy-scale uncertainty, assigned a prior of $1\%$; 
$\delta_2$ parametrizes the uncertainty in the energy-resolution model, with a prior 
set to $5\%$; and $\delta_3$ describes non-linearity effects in the reconstructed energy scale, for which a prior uncertainty of $1\%$ is adopted. 

The predicted number of events in the $i$-th prompt energy bin is
\[ 
  N_i(\alpha,\beta,\gamma)
= (1+\alpha)\,S_i\,[1 + \gamma(E_i - \langle E\rangle)] + (1+\beta)\,B_i ,
\]
where $B_i$ is the total background spectrum in that bin.

The $\chi^2$ function used to extract constraints is
\begin{equation}
\begin{aligned}
\chi^2 = \min_{\alpha,\beta,\gamma,\{\delta_k\}} \bigg[
& \sum_i \frac{(N_i^{\rm meas}-N_i)^2}{\sigma_{i,\rm{stat}}^2}
  + \left(\frac{\alpha}{\sigma_{\alpha}}\right)^2  \\
& + \left(\frac{\beta}{\sigma_{\beta}}\right)^2
  + \left(\frac{\gamma}{\sigma_{\gamma}}\right)^2
  + \sum_{k=1}^{3} \left( \frac{\delta_k}{\sigma_{\delta_k}} \right)^2
\bigg],
\end{aligned}
\label{eq:chisquared}
\end{equation}
where $N_i^{\rm meas}$ is the observed number of events in bin $i$, 
$\sigma_{i,\rm stat}=\sqrt{N_i^{\rm meas}}$, and 
$\sigma_\alpha$, $\sigma_\beta$, $\sigma_\gamma$, and $\sigma_{\delta_k}$ for $k=1,2,3$ are the priors uncertainties associated with the corresponding nuisance parameters.

The following sections present the results on light sterile neutrinos and scalar NSI.

\section{Constraints on Light Sterile Neutrinos}

\begin{figure}[htbp!]
    \centering
    \includegraphics[width=\linewidth]{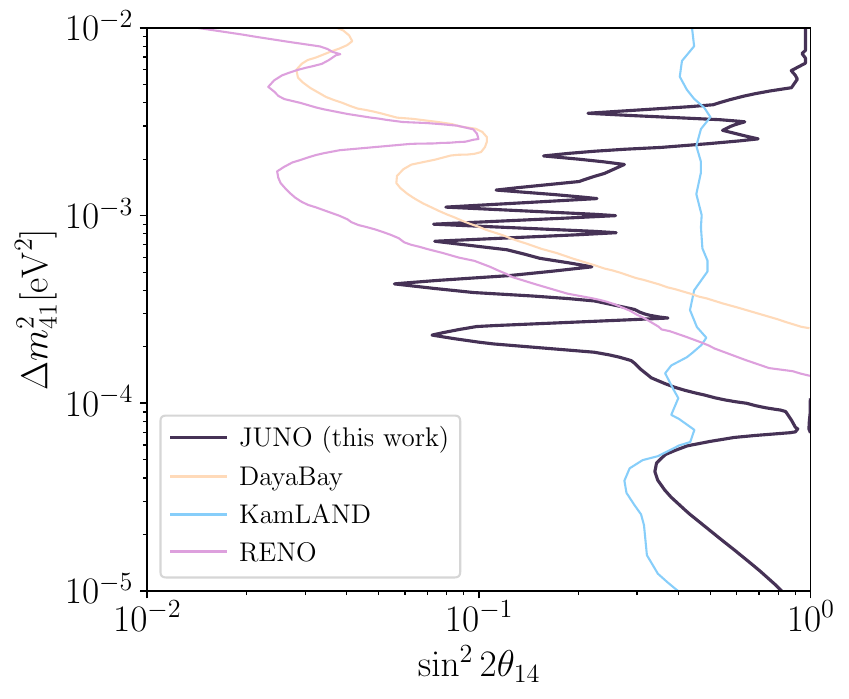}
\caption{95\% CL exclusion contour for the neutrino oscillation parameters $\sin^2 \theta_{14}$ and $\Delta m^2_{41}$, assuming normal mass ordering. The constraints obtained from the JUNO data analyzed in this work are presented (black), as well as limits from KamLAND~\cite{Chen2022}, Daya Bay~\cite{An2014}, and RENO~\cite{Choi2020}.}
    \label{fig:theta_deltam_st}
\end{figure}

The resulting constraints from the JUNO data analyzed in this work are shown in Fig.~\ref{fig:theta_deltam_st}, assuming normal mass ordering. The analysis probes the light sterile-neutrino parameter space in the region $10^{-5} \,\mathrm{eV}^2\lesssim \Delta m_{41}^2 \lesssim 10^{-2}\,\mathrm{eV}^2$, reaching sensitivity to mixing values of $\sin^2 2\theta_{14}$ at the level of $10^{-1}$. These results already probe previously unexplored parameter space in the range $\Delta m^2_{41} \approx (1$--$5)\times 10^{-4}$~eV$^2$. Bounds from KamLAND, Daya Bay, and RENO are also shown for comparison.

\section{Constraints on Scalar NSI Parameters}

\begin{figure}[htbp!]
    \centering
    \includegraphics[width=\linewidth]{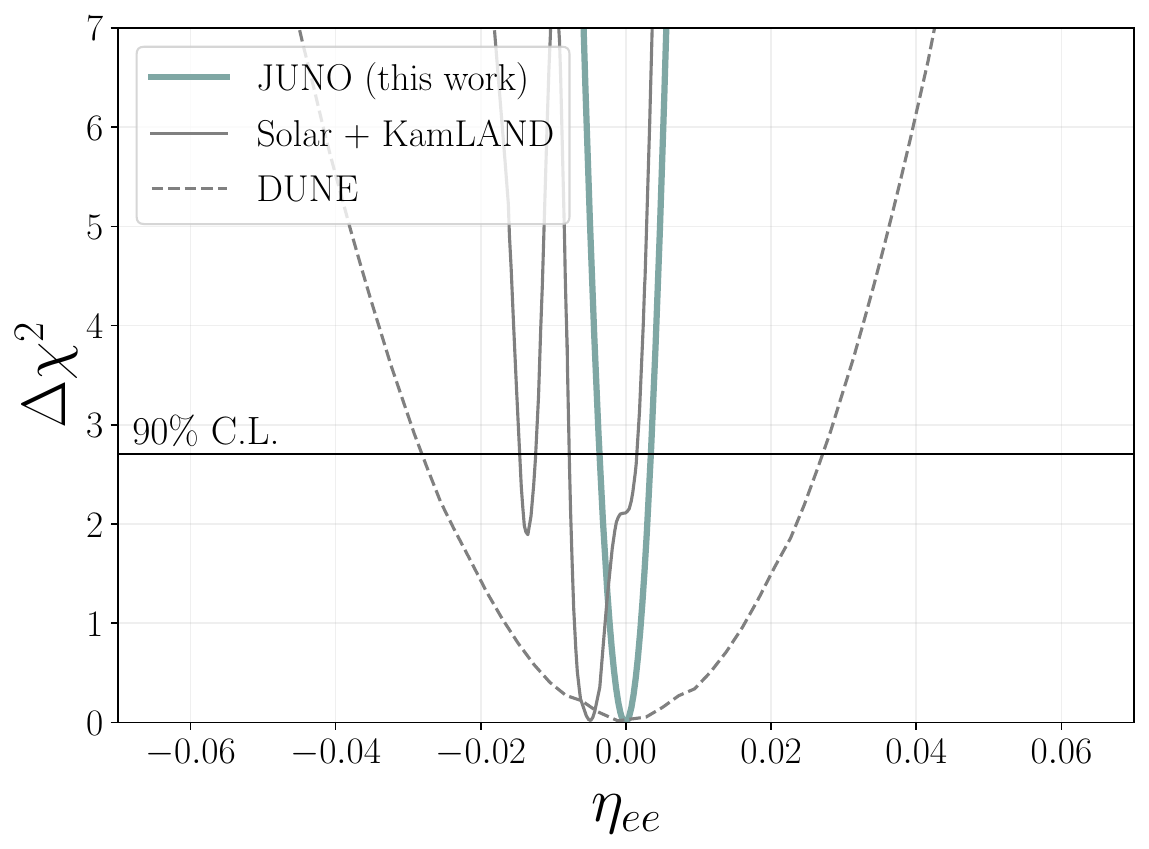}
    \caption{$\Delta\chi^2$ profile of the scalar $\eta_{ee}$ parameter obtained in this work. Results from a combination of solar and KamLAND data~\cite{Denton:2024upc} are also shown, together with the expected sensitivity from DUNE~\cite{Singha:2023}.}
    \label{fig:etaee}
\end{figure}

The constraints on the scalar NSI parameter $\eta_{ee}$ obtained from the JUNO data analyzed in this work are shown in Fig.~\ref{fig:etaee}. The analysis yields limits at the level of $-0.0036 < \eta_{ee} < 0.0034$ at 90\% C.L. These results demonstrate that reactor antineutrino spectral measurements provide complementary sensitivity to scalar NSI. For comparison, results from a combined solar and KamLAND analysis~\footnote{ In the present convention, $\delta M=\sqrt{|\Delta m^2_{31}|}\,\eta$, while Ref.~\cite{Denton:2024upc} uses $\delta M(r)=\sqrt{\Delta m^2_{21}}\,(\rho(r)/\rho_\odot)\,\eta^\odot$, with $\rho_\odot=100\,\mathrm{g/cm^3}$. Thus, a direct comparison requires a normalization change and a density rescaling, giving $\eta^\odot=(\sqrt{|\Delta m^2_{31}|}/\sqrt{\Delta m^2_{21}})(\rho_\odot/\rho_{\rm JUNO})\,\eta$, with $\rho_{\rm JUNO} = 2.6\,\mathrm{g/cm^3}$. }~\cite{Denton:2024upc} are also shown, which constrain the parameter to the range $-0.0078 < \eta_{ee} < 0.0015$ at 90\% C.L., along with the expected sensitivity from DUNE~\cite{Singha:2023}.

\begin{figure}[htbp!]
    \centering
    \includegraphics[width=\linewidth]{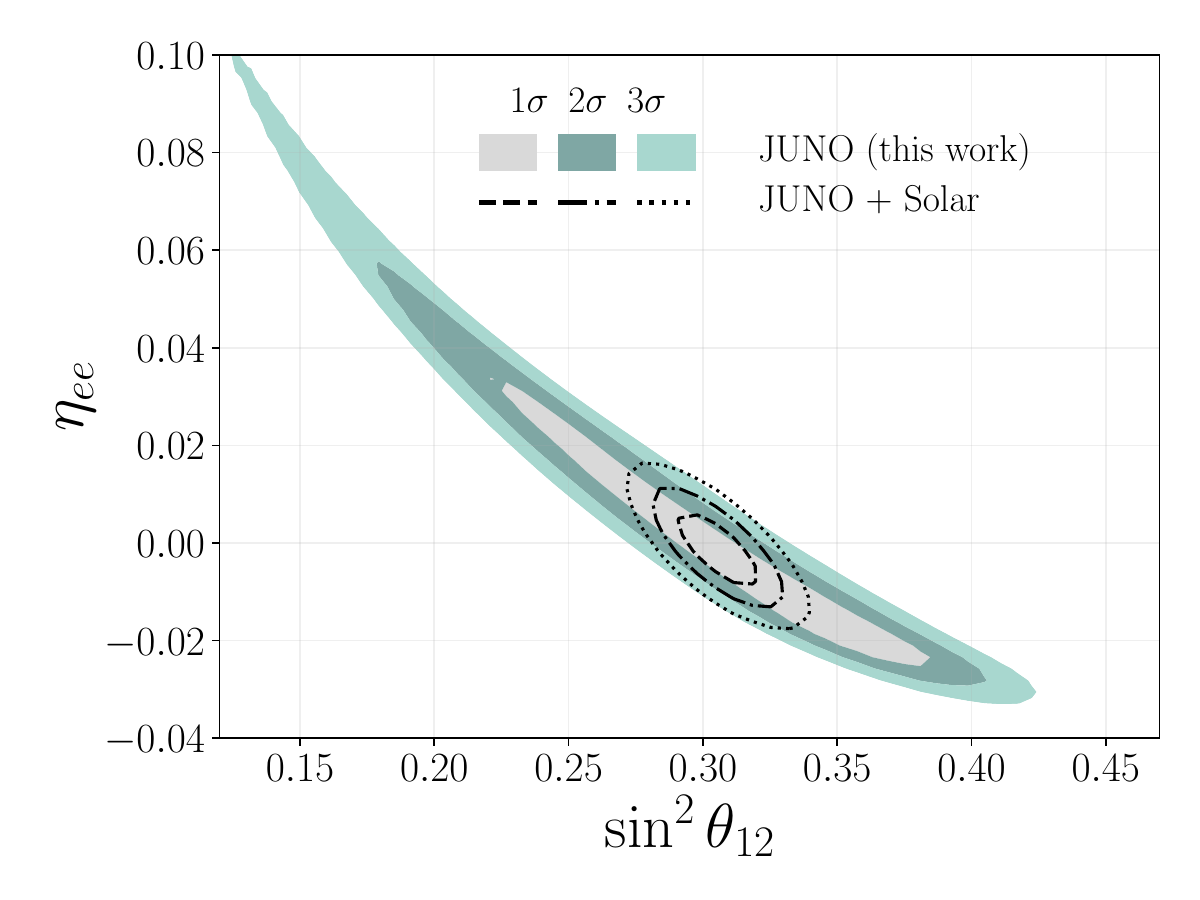}
    \caption{Allowed region in the $\sin^{2}\theta_{12}$--$\eta_{ee}$ parameter space. The filled contour corresponds to the JUNO-only fit performed in this work, while the line contours include the solar prior on $\sin^2\theta_{12}$ and $\Delta m^2_{21}$~\cite{Super-Kamiokande:2023jbt}. Contours indicate the $1\sigma$, $2\sigma$, and $3\sigma$ confidence regions.}
    \label{fig:nsi_s12}
\end{figure}

\begin{figure}[htbp!]
    \centering
    \includegraphics[width=\linewidth]{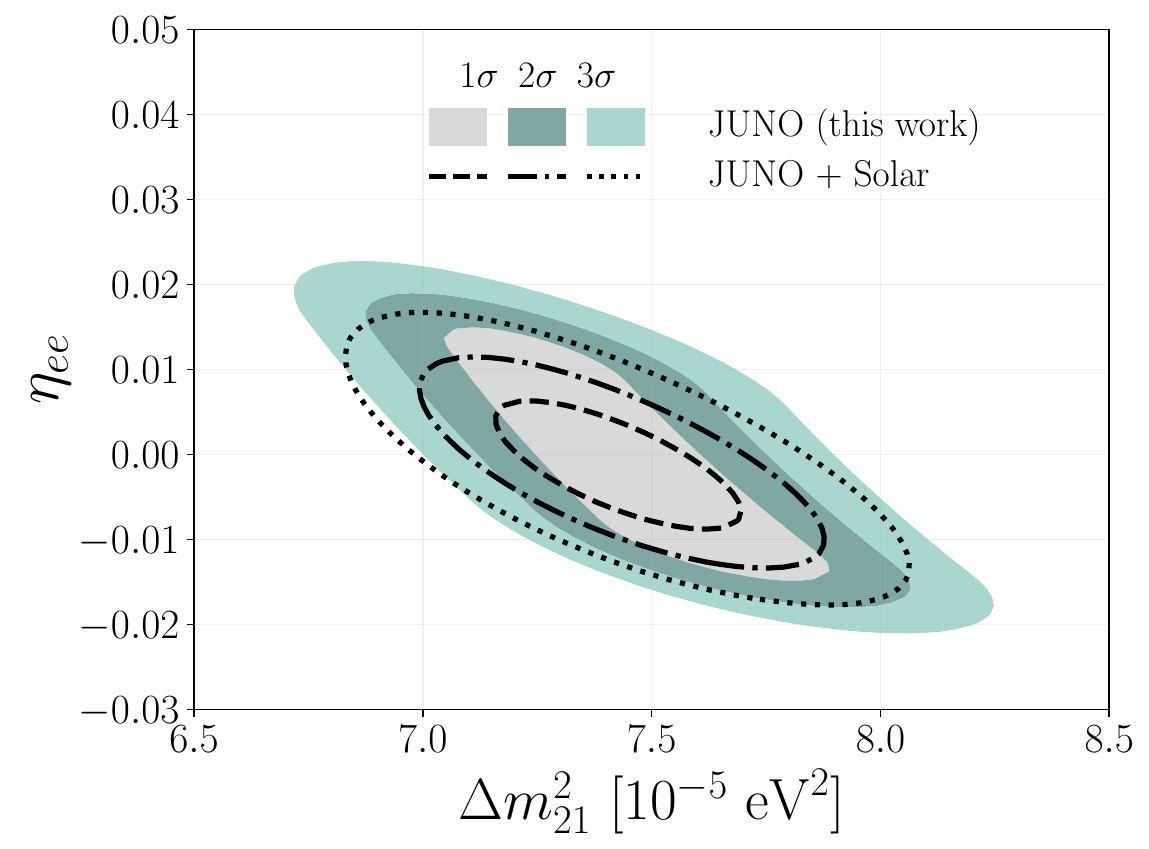}
    \caption{Allowed region in the $\Delta m^{2}_{21}$--$\eta_{ee}$ parameter space. The filled contour corresponds to the JUNO-only fit performed in this work, while the line contours include the solar prior on $\sin^2\theta_{12}$ and $\Delta m^2_{21}$~\cite{Super-Kamiokande:2023jbt}. Contours indicate the $1\sigma$, $2\sigma$, and $3\sigma$ confidence regions.}
    \label{fig:nsi_dm21}
\end{figure}

The results also show correlations between $\eta_{ee}$ and the solar oscillation parameters $\sin^2\theta_{12}$ and $\Delta m_{21}^2$, as shown by the allowed regions in Figs.~\ref{fig:nsi_s12} and \ref{fig:nsi_dm21}. These correlations originate from modifications to the effective neutrino Hamiltonian induced by scalar interactions and illustrate the capability of precision reactor experiments to probe such effects. In addition to the JUNO-only fit, a combined fit is performed  incorporating constraints from solar neutrino experiments. This is implemented by adding Gaussian prior terms on the parameters $\sin^2\theta_{12}$ and $\Delta m^2_{21}$ to the $\chi^2$ function in Eq.~\eqref{eq:chisquared}. The central values and uncertainties are taken from global analyses of solar neutrino data~\cite{Super-Kamiokande:2023jbt}. These priors help to partially break the degeneracy between the scalar NSI parameter $\eta_{ee}$ and the solar oscillation parameters $\sin^2\theta_{12}$ and $\Delta m^2_{21}$.

Since scalar NSI effects scale with the matter density, experiments probing neutrino propagation in dense environments, such as solar neutrino experiments, are expected to provide stronger constraints on these interactions~\cite{Denton:2024upc}. The results obtained here provide complementary sensitivity to scalar NSI in a terrestrial low-density regime characterized by distinct parameter degeneracies, in addition to different systematic uncertainties across experiments.

\section{Conclusions}

The results of this work show that the first JUNO reactor antineutrino data already provide meaningful sensitivity to oscillatory features induced by light sterile neutrinos and to subleading deviations associated with scalar NSI effects. The parameter region explored for light sterile neutrinos is complementary to that probed by short-baseline reactor experiments and accelerator-based searches, emphasizing the importance of JUNO in covering previously inaccessible regions of parameter space. In the case of scalar NSI, degeneracies with standard oscillation parameters limit the capability of a single experiment to fully disentangle these effects, underscoring the need for complementary measurements from reactor, solar, and accelerator neutrino experiments. While solar neutrino experiments currently provide the most stringent constraints on scalar NSI due to the higher matter densities involved, the JUNO results presented here offer a complementary probe in a distinct density environment, with different parameter degeneracies, as well as different systematic uncertainties. The constraints obtained in this analysis are expected to improve significantly as JUNO accumulates additional exposure and achieves better control of systematic uncertainties.

\section{Acknowledgements}
\begin{acknowledgments}
\indent This work is supported by SECIHTI Project No. CBF-2025-I-1589 and DGAPA UNAM Grants No. PAPIIT IN102326, and IN111625. GSG was supported by the post-doctoral fellowship program DGAPA-UNAM. The authors thank Peter B. Denton for useful comments regarding the interpretation of scalar NSI constraints and their comparison with solar neutrino results.

\end{acknowledgments}
\sloppy
\hbadness=10000
\bibliographystyle{apsrev4-1}
\bibliography{bibliography}

\end{document}